# ADIABATIC FOCUSER WITH ION MOTION

E. Henestroza, A.M. Sessler, S. S. Yu, LBNL, Berkeley, California, U.S.A.*

*Abstract*

In this paper, after reviewing the concept of an adiabatic focuser we numerically study the effect of ion motion. Our work has been motivated by a recent suggestion that ion motion in an adiabatic focuser might be significant and even preclude operation of the focuser as previously envisioned. It is shown that for parameters associated with cases in which the adiabatic lens would be used, despite ion motion, the adiabatic focuser should work as well as originally envisioned. We then extend the study by considering a very long electron pulse (which accentuates the effect of ion motion).

## INTRODUCTION

The adiabatic focuser [1] works by having a focusing channel whose strength increases with distance down the channel. In this situation, electrons of various energies and various transverse oscillation phase all are transversely focused. The concept works with external focusing, but would be very effective in a plasma ion- focusing channel where the density of ions is simply increased as one goes down the channel. In the original work [1] motion of the ions was not included (as it was assumed to be a small effect).

Recently, it has been suggested that ion motion in an adiabatic focuser might be significant and possibly preclude operation of the focuser as previously envisioned [2]. In order to respond to this concern, in this paper we remove the previously made assumption that the ions do not move and numerically study ion motion in the adiabatic focuser. The ions clearly influence each other and, most importantly, are influenced by the electric field of the electrons being focused. However, taking parameters as originally developed, it is shown that, despite ion motion, the adiabatic focuser works as well as originally envisioned. We then examine very long electron bunches and show that ion motion degrades the performance of the adiabatic focuser. Thus a criterion is developed for bunch length such that ion motion is negligible.

In the following we first (Section II) review the few equations that characterize the adiabatic focuser. This brief review serves to define the notation used in this paper. In Section III we describe the computational model employed, and in Section IV the results of the numerical studies are presented. This report is an extension of the work presented at the EPAC2006 conference [3], by presenting more details of the earlier studies and most particularly by studying, carefully, the consequence of a very long electron bunch so that the effect of ion motion is significant .

## ADIABATIC FOCUSER

In Ref. 1, the adiabatic focuser is developed using a 1D analysis appropriate, for example, to a flat beam. Although the numerical analysis in this paper is 3D, we follow the previous analysis, which is adequate to define terms. Let $y$ be the transverse coordinate and $s$ the distance along which the beam travels. The equation of motion is

$$\frac{d^2 y}{ds^2} + K(s) y = 0, \tag{1}$$

with solution

$$y(s) = \beta^{1/2}(s) \cos[\psi(s) + \varphi], \tag{2}$$

where

$$\frac{d\beta}{ds} = -2\alpha(s), \tag{3}$$

* Work supported by the U.S. Department of Energy, Office of Basic Energy Sciences, under Contract No. DE-AC02-05CH11231

and

$$\psi(s) = \int^s \frac{ds'}{\beta(s')}. \tag{4}$$

We have explored various choices for the function $\beta(s)$, and found that the adiabatic focuser works for all choices, and the conclusions of this paper are also essentially independent of this choice. For simplicity we take $d\beta/ds$ constant and, hence,

$$\beta(s) = \beta_0 - 2\alpha_0 s, \tag{5}$$

where $\beta_0$ is the initial value of $\beta$ while $\alpha_0$ is the initial condition (and a constant throughout the focuser) and a measure of the degree of adiabaticity of the focuser.

The strength of the channel is readily determined as

$$K(s) = \frac{1+\alpha_0^2}{\beta(s)^2} = \frac{1+\alpha_0^2}{(\beta_0 - 2\alpha_0 s)^2}. \tag{6}$$

If we make the focusing channel from a plasma, this formula tells us how to vary the ion density from its initial value of $n_0$ to its final value, obtained in length $L$, of $n^*$.

The high-energy beam is characterized by its energy $E_0$, its normalized emittance $\varepsilon_n$ and spot size $\sigma (=\sqrt{\beta\varepsilon})$ where $\varepsilon$ is the true emittance, i.e., $\varepsilon_n/\gamma$. The initial beam spot size is $\sigma_0$ and the final spot size is $\sigma^*$.

## EFFECT OF ION MOTION

The effect of ion motion is complicated and is by necessity addressed numerically in the rest of this paper. The ions move under the influence of the electric field created both by the high-energy electrons and the ions. Clearly only ion motion within the region of the electron beam will affect the electrons. Thus ions that move into that region and also ions within that region that change their position are both of concern.

We can get a rough determination of the magnitude of the effect by focusing upon the ions within the beam. (Those moving in to the beam are subject to one field outside and a different field inside, so the calculation is more complex, but readily do-able. The parametric behavior is the same as derived below.) Taking the beam as a cylinder of large axial extent (and since the compression effect is adiabatic this is a good approximation) we can readily compute the electric field as:

$$E_r = -\frac{\lambda e}{2\pi\varepsilon_0 a^2} r, \tag{7}$$

where $\lambda$ is the line number density. There are electric fields due to both ions and high energy electrons. In the underdense case, the field from the electrons dominates and so $\lambda$ can be related to the electron beam current by $I = \lambda c$.

The equation of motion of the ions, of mass $M$, moving under this field is given by:

$$M\frac{d^2 r}{dt^2} = eE_r, \tag{8}$$

The ions therefore perform harmonic oscillations with period $\tau$ and characteristic beam displacement $\ell$ given by:

$$\ell = c\tau = a\sqrt{\frac{2\pi^2 M}{r_0 \lambda m}}, \tag{9}$$

where, $m$ is the electron mass, $a$ is the electron beam radius, $r_0$ is the classical electron radius, and $\ell$ is the length of electron bunch beyond which ion motion effects would be important. Inserting the parameters of the numerical work, gives a result in the *mm* range for densities up to *5e18 cm$^{-3}$*, which is quite consistent with the results displayed in Fig 9.

# COMPUTATIONAL MODEL

The numerical studies were carried out using the code WARP [4]. The code was used to study the motion of the high-energy electrons and the ions. The low-energy electrons (resulting from the ionization process) were not included in the calculation. Although WARP is capable, in principle, of calculating the beam dynamics, including both the secondary (slow) electrons and the ions, we adopt for this paper a simple approximation based on ion motion alone. The approximation is based on the physical observation that secondary electrons are instantly expelled with the entry of the high-energy electron bunch. This secondary electron "clearing" will take place at any radial position as long as the charge of the ions enclosed is less than the charge of the high-energy beam. The secondary electrons will "pile up" at a radial position where the net charge enclosed is zero. In the "underdense" regime considered in this paper, the secondary-electron pile-up will take place outside of the high-energy bunch. By Gauss' Law, the secondary electrons will exert no force on the high-energy bunch, and can be ignored. The ions can then be modelled by a "bare" column with ion charge equal in magnitude to the charge of the high-energy bunch.

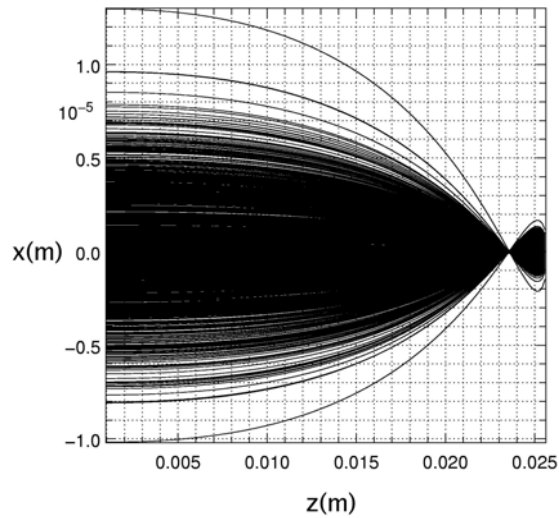

Figure 1: Electron trajectories with the ions fixed, and the high-energy electron bunch having zero emittance. The ions supply a focusing force as prescribed by Eq. 6.

The focuser works by first establishing a focusing channel of increasing strength, probably by means of differential pumping and then firing in an ionizing electron beam. The resulting plasma of low-energy electrons and ions remains (almost) fixed as it is electrically neutral. Then, subsequently, the high-energy pulse of electrons is fired into the plasma.

To model this system without including the low-energy electrons, we adopt the following method. First a plasma distribution is constructed with increasing density down the channel according to the prescription of the adiabatic focuser. Then, as the high-energy electrons enter the plasma they eject radially the plasma electrons in a short time compared to the electron pulse length, and create ions distributed radially over a region such that the number of ions in that region are equal to the number of beam electrons. Since the electron density is always higher than the ion density, one appreciates that the low-energy electrons are pushed out by the high-energy electrons until charge neutrality is obtained, i.e., over a larger radial region than the electron beam. This process is then repeated at each step. Once the ions are "created" they are free to move. (They don't move prior to the high-energy beam entering the plasma due to the presence of the low-energy electrons.) The number of macroparticles representing the electrons injected per time step is 5000 and the number of macroparticles representing the plasma ions that appear after ejecting the low energy electrons is, for convienence, also equal to 5000. The transverse grid size is 4e-6 cm and we have used 250 cells to cover an extension of 10 μm. The longitudinal grid size is 0.1 mm and we have used 256 cells to simulate the whole adiabatic focuser of length of 2.56 cm. The time step is chosen to push each set of injected beam electrons by a distance of one grid size ($\Delta t = \Delta z/c$).

# RESULTS

## A. Nominal Parameters

The results of the numerical studies are presented in the figures. In the numerical computation we employed the parameters exhibited in Table I. At first we considered an "ideal adiabatic lens", that is, one where the ions are in fixed locations and the high energy electrons have zero emittance. The result is shown in Fig. 1 and will provide a basis for comparisons with the more complicated cases next studied.

Table 1: Parameters of the adiabatic focuser studied numerically

| Initial Beam Parameters | |
|---|---|
| $E_0$ (GeV) | 50 |
| $\ell$ (mm) | 1 |
| $\varepsilon_n$ ($\pi$-m-rad) | $3 \times 10^{-5}$ |
| $\sigma_0$ ($\mu$m) | 3 |
| $\beta_0$ (cm) | 3 |
| **Focuser Properties** | |
| $\alpha_0$ | $1/\sqrt{3}$ |
| L (cm) | 2.6 |
| $n_0$ (cm$^{-3}$) | $8.4 \times 10^{15}$ |
| $n^*$ (cm$^{-3}$) | $8.4 \times 10^{19}$ |
| **Final Beam Property** | |
| $\sigma^*$ ($\mu$m) No ion motion | 0.3 |
| $\sigma^*$ ($\mu$m) With ion motion | 0.3 |

The first new result, namely, one in which the ions are allowed to move, is presented in Fig. 2. The electron beam is taken to have zero emittance so that a comparison with our base case (Fig. 1) can readily be made. The focuser is seen to be as effective with ion motion included as it was when we assumed the ions were fixed. At the electron beam head, the ions have no time to move, and the beam envelope is therefore identical to the frozen-ion case. Subsequent electron-beam slices see increasingly more ion bunching. Both the ion number as well as the radial distribution will change with time at a fixed axial location. However, a given electron slice will see an adiabatically varying ion distribution (with z), and the concept of adiabatic focusing will continue to be effective.

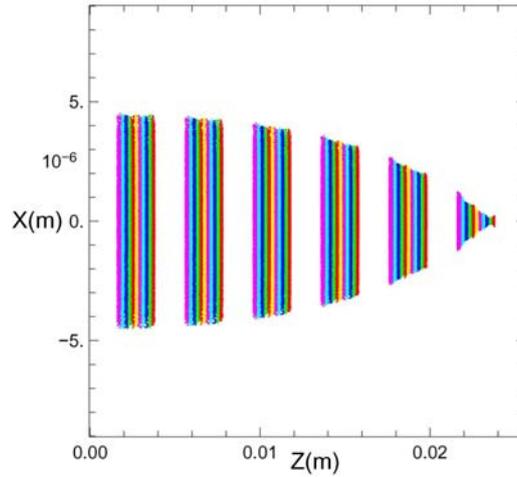

Figure 2: Six snap shots of a section (1 mm back from the head) of the high-energy electron bunch as it moves through the adiabatic lens focuser. The ions are allowed to move, and the high-energy electron bunch has zero emittance.

In Fig. 3 we give the electron beam non-zero emittance and, once again, the focuser is seen to be effective.. In fact, there is no significant different between Figure 2 and Figure 3.

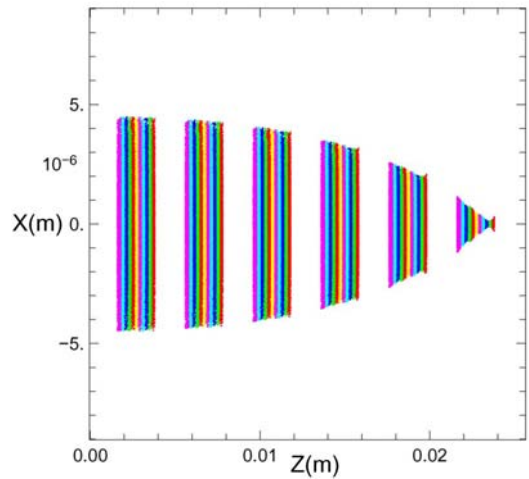

Figure 3: Six snap shots of a section (1 mm back from the head) of the high-energy electron bunch as it moves through the adiabatic lens focuser. The ions are allowed to move, and the high-energy electron bunch has non-zero emittance.

Finally, in Figs. 4 and 5 we present some typical trajectories so that the reader can see, in detail, the strong focusing of electrons and the modest (inward) motion of the ions.

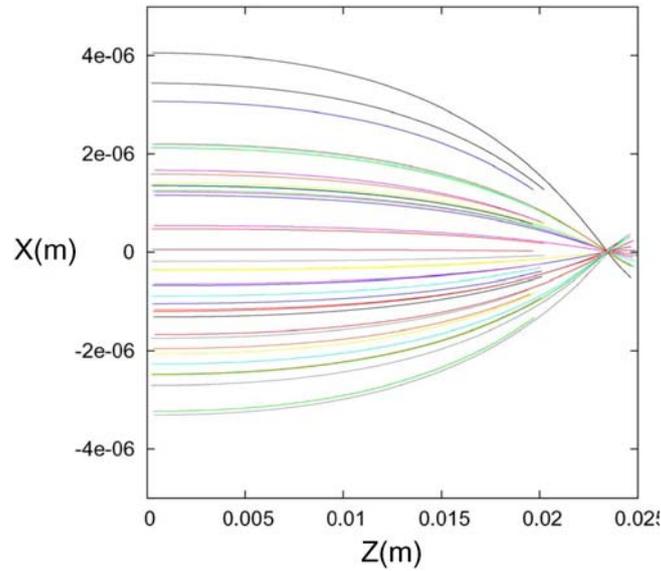

Figure 4: A few typical electron trajectories in the adiabatic focuser are shown.

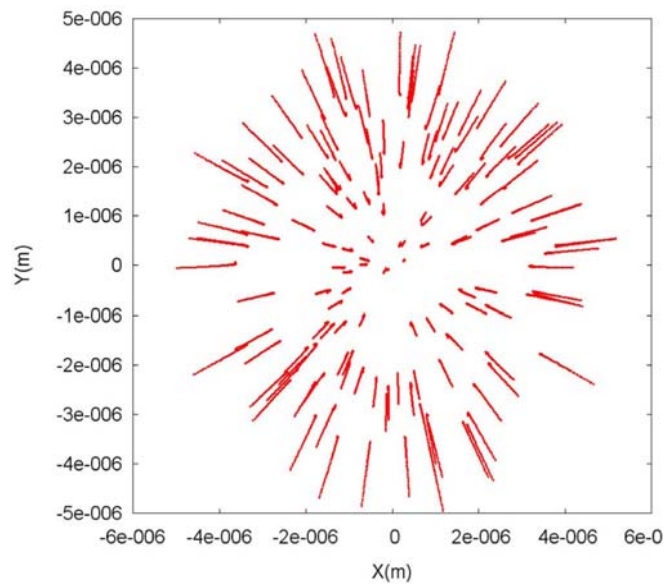

Figure 5: A few typical ion trajectories, at a fixed z location, in the adiabatic focuser are shown. One can see how the ions are first pulled in (focused) by the high-energy electrons and then blown up after the bunch passes by.

## B. Long Beam Bunch

As the bunch length increases we expect that the effect of ion motion will become greater. That is, as the ions have time enough to move significantly (the consequence of a long bunch) the focusing strength will be changed from that expected with non-moving ions. Thus the rear of the electron bunch will feel a different focusing force than that felt by a particle at the front of the bunch. The result will be that the particles will come to a focus at a different longitudinal location. Hence the focal spot of the beam will be not as good as in the non-moving ion case. In order to study this effect, we simulated a long bunch going through the adiabatic focuser. Fig. 6 shows the ion profiles at an axial location 16 mm upstream of the focal plane at the time (left) when the beam head expels the slow electrons from the plasma, and the density profile (right) 30 picoseconds afterwards. The ions are being focused by the high-energy electron beam as shown by the high density at the center of the ion cloud.

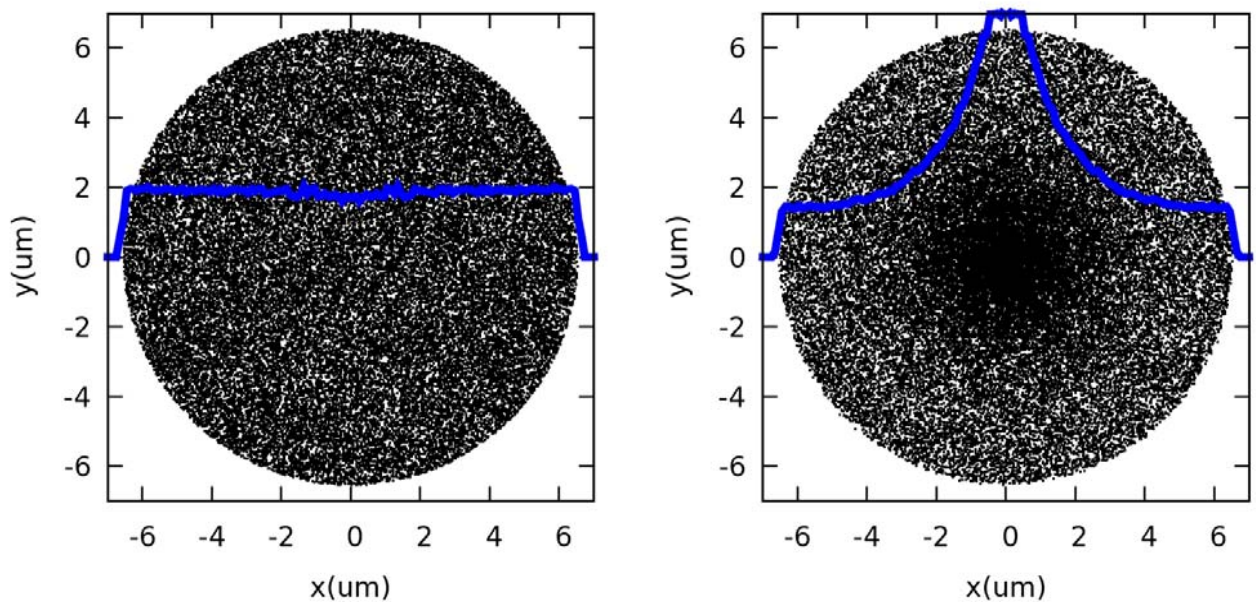

Figure 6: Ion cloud profile and central linout at a fixed z location, in the adiabatic focuser at two different times. On the left is the uniform density profile as generated by the high-energy electron beam head. On the right is the non-uniform density profile 30 ps afterwards.

In Fig 7 we display the effect of ion motion on the focal point produced by the adiabatic focuser. As can be seen different parts of a long bunch come to focus at different positions and the degree of focus also varies throughout the bunch.

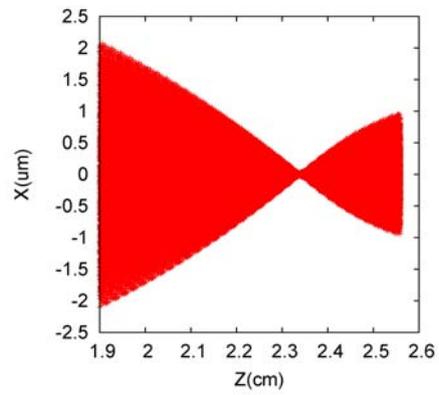

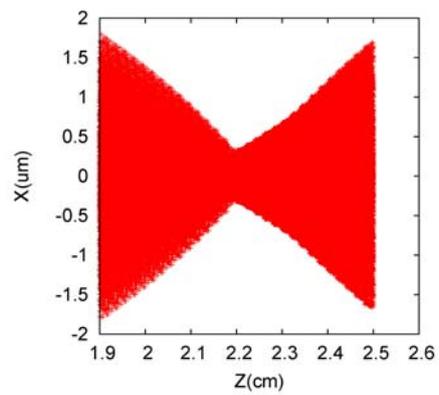

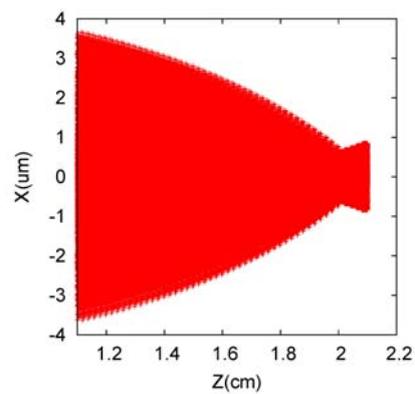

Figure 7: Profile of the motion of three electron bunch slices at 1, 5 and 10 mm from the head (top to bottom), as each slice moves through the adiabatic lens focuser. The head is well focused and the subsequent slices focus less due to the ion motion.

The effect of ion motion on the rms-radius of the beam is shown in Fig. 8 as a function of arrival time at several axial locations. The effect of ion motion is seen by the increase in spot size about 10 picoseconds after the passage of the electron beam head. However, for nominal parameters the bunch length is around a few picoseconds (Table 1), one can conclude that the ion motion does not preclude the final focusing of the electron bunch to a small spot by the adiabatic focuser.

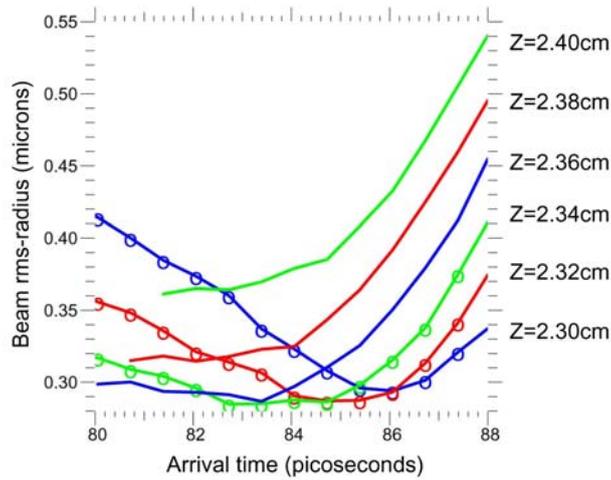

Figure 8: Plots of beam radius (rms) as function of arrival time at several fixed z locations in the adiabatic focuser.

The average spot size as a function of bunch length (at z = 2.36 cm) taken from the curve in Fig. 8. is shown in Fig 9. One sees for short bunches the effectiveness of the adiabatic lens, but as the bunch becomes longer the lens is less effective due to the ion motion.

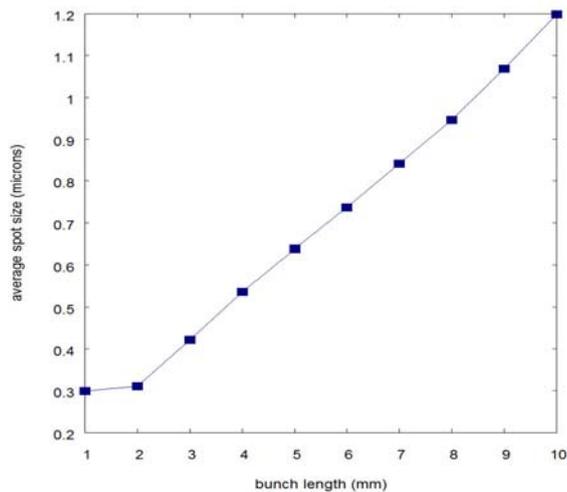

Figure 9: Average spot size of an electron bunch as a function of the bunch length.

## CONCLUSIONS

The numerical studies were first carried out for the adiabatic focuser whose parameters are exhibited in Table I. We have shown that even when the bunch length is considerably increased, up to certain limit over the nominal value, the ion motion is still a small effect upon the ability of the adiabatic focuser to produce a small focal spot. It is clear however, since the high-energy electrons simply focus the ions even more than if they are held fixed, and reflecting upon how the adiabatic focuser "works", it is evident that the general conclusion that ion motion will not preclude the effective operation of an adiabatic focuser, is true in general. We should also note that the ion motion can be reduced by going to heavier gases, as long as the electron-beam scattering effects are insignificant.

A second series of numerical studies involved a very long bunch. In fact we have considered bunches long enough that ion motion is significant. We have thus determined the length of a bunch at which ion motion degrades the adiabatic lens to a significant degree as is shown in our final Fig. 9.